\documentclass[a4paper,fleqn,usenatbib]{mnras}

\usepackage[T1]{fontenc}
\usepackage{ae,aecompl}

\usepackage{graphicx}	
\usepackage{amsmath}	
\usepackage{amssymb}	
\usepackage{multirow}
\usepackage{booktabs}
\usepackage{caption}
\usepackage{natbib,hyperref}

\newcommand{\zR}{$\zeta^2\,\mathrm{Reticuli}\,$}
\newcommand{\excs}{\extracolsep{\fill}}

\title[Is there really a debris disc around \zR ?]{Is there really a debris disc around \zR ?}

\author[Faramaz et al.]{
 V. Faramaz,$^1$\thanks{E-mail: virginie.c.faramaz@jpl.nasa.gov} 
  G. Bryden,$^1$
 K. R. Stapelfeldt,$^1$ 
 M. Booth,$^2$
  A. Bayo,$^{3,4}$ 
 H. Beust,$^5$ 
 \newauthor
 S. Casassus,$^{6,7}$ 
 J. Cuadra,$^{8,4}$ 
 A. Hales,$^{9,10}$ 
 A. M. Hughes,$^{11}$ 
 J. Olofsson,$^{3,4}$
  K. Y. L. Su,$^{12}$
  \newauthor
  D. J. Wilner$^{13}$
\\
$^1$ Jet Propulsion Laboratory, California Institute of Technology, 4800 Oak Grove drive, Pasadena CA 91109, USA.\\
$^2$ Astrophysikalisches Institut und Universit{\"a}tssternwarte, Friedrich-Schiller-Universit{\"a}t Jena, Schillerg{\"a}{\ss}chen 2-3, 07745 Jena, Germany.\\
$^3$ Instituto de F\'isica y Astronom\'ia, Facultad de Ciencias, Universidad de Valpara\'iso, Av. Gran Breta\~na 1111, Valpara\'iso, Chile.\\
$^4$ N\'ucleo Milenio Formaci\'on Planetaria - NPF, Chile \\
$^5$ Univ. Grenoble Alpes, CNRS, IPAG, F-38000 Grenoble, France.\\
$^6$ Departamento de Astronomia, Universidad de Chile, Casilla 36-D, Santiago, Chile.\\
$^7$ Millennium Nucleus "Protoplanetary discs", Santiago, Chile.\\
$^8$ Instituto de Astrof\'isica, Facultad de F\'isica, Pontificia Universidad Cat\'olica de Chile, 782-0436 Santiago, Chile.\\
$^9$ Joint ALMA Observatory, Alonso de C\'ordova 3107, Vitacura 763-0355, Santiago, Chile.\\
$^{10}$ National Radio Astronomy Observatory, 520 Edgemont Road, Charlottesville, Virginia, 22903-2475, USA.\\
$^{11}$ Department of Astronomy, Van Vleck Observatory, Wesleyan University, Middletown, CT 06459, USA.\\
$^{12}$ Steward Observatory, University of Arizona, 933 N Cherry Ave., Tucson, AZ 85721, USA.\\
$^{13}$ Harvard-Smithsonian Center for Astrophysics, 60 Garden Street, Cambridge, MA 02138, USA.
}

\date{Accepted 2018 August 21. Received 2018 August 20; in original form 2018 June 20.}

\pubyear{2018}

\begin{document}
\label{firstpage}
\pagerange{\pageref{firstpage}--\pageref{lastpage}}
\maketitle

\begin{abstract}
The presence of a debris disc around the Gyr-old solar-type star \zR was suggested by the \textit{Spitzer} infrared excess detection. Follow-up observations with \textit{Herschel}/PACS revealed a double-lobed feature, that displayed asymmetries both in brightness and position. Therefore, the disc was thought to be edge-on and significantly eccentric. Here we present ALMA/ACA observations in Band 6 and 7 which unambiguously reveal that these lobes show no common proper motion with \zR. In these observations, no flux has been detected around \zR that exceeds the $3\sigma$ levels. We conclude that surface brightness upper limits of a debris disc around \zR are $5.7\,\mathrm{\mu Jy/arcsec^2}$ at 1.3 mm, and $26\,\mathrm{\mu Jy/arcsec^2}$ at 870 microns. Our results overall demonstrate the capability of the ALMA/ACA to follow-up \textit{Herschel} observations of debris discs and clarify the effects of background confusion.
\end{abstract}



\section{Introduction}\label{sec:intro}

\zR (HR 1010, HIP 15371, HD 20807) is a 3 Gyr-old G1V solar-type star \citep{2013A&A...555A..11E,2014ApJ...785...33S}, located at 12 pc \citep{2007A&A...474..653V,2016A&A...595A...1G,2018arXiv180409365G}. The presence of a debris disc surrounding \zR was first suggested by observations carried out with \textit{Spitzer}/MIPS \citep{2004ApJS..154...25R}, which detected a 5\% infrared excess of this star compared to the predicted photospheric flux at 24 microns, as well as a more significant, 77\% infrared excess at 70 microns \citep{2008ApJ...674.1086T}. 
\zR was later observed with \textit{Herschel}/PACS \citep{2010A&A...518L...2P} and SPIRE \citep{2010A&A...518L...3G}, in the frame of the \textit{Herschel} Open Time Key Programme DUst around NEarby Stars (DUNES). While \textit{Herschel}/SPIRE's resolution is insufficient to discern much structural detail, that of \textit{Herschel}/PACS at 70 and 100 microns, $\sim 6-7$", allowed the resolution of a double-lobed structure at these wavelengths. The lobes are seen asymetrically distributed on each side of the star, with the lobe closest to the star (South-East) being brighter than its North-West counterpart \citep{2010A&A...518L.131E,2014A&A...563A..72F}.

This feature was interpreted as a debris disc extending from $\sim 70\,$AU (6") to $\sim 120\,$AU (10"), seen close to edge-on and significantly eccentric $(e\geq 0.3)$. This interpretation was supported by several arguments. 

\newpage

\begin{enumerate}
\item The inclination of the star's rotation axis with respect to the line of sight, which was found to be $65^{\circ\,+24.5}_{-31.5}$ \citep[Appendix B]{2014A&A...563A..72F}. Indeed, since observations suggest that debris disc orientations differ little from that of their host star \citep{2011MNRAS.413L..71W,2014MNRAS.438L..31G} a debris disc surrounding \zR would be expected to be seen close to edge-on.
\item The position asymmetry of the lobes. Indeed, stars surrounded by an eccentric debris disc are offset from the disc centre of symmetry, as exemplified by the debris discs of Fomalhaut \citep{2005Natur.435.1067K,2005ApJ...620L..47M}, HD 202628 \citep[][Faramaz et al., in prep.]{2012AJ....144...45K}, and HR4796 \citep{2009AJ....137...53S,2011ApJ...743L...6T}.
\item The brightness asymmetry. At (far) infrared wavelengths, eccentric debris discs are expected to show a "pericentre-glow" at the inferred position of periastron \citep{1999ApJ...527..918W}, as material closer to the star is expected to be hotter and brighter, again exemplified by observations of the debris disc of Fomalhaut \citep{2004ApJS..154..458S,2012A&A...540A.125A}, HD 202628 (Faramaz et al., in prep.), and HR 4796 \citep{2000ApJ...530..329T,2011A&A...526A..34M}.
\end{enumerate}

At longer (sub)mm wavelengths, the brightness asymmetry of such eccentric debris discs is expected to reverse, with the apocenter being this time brighter than the pericenter. This phenomenon, called "apocentre-glow" \citep{2016ApJ...832...81P}, is due to the fact that eccentric debris discs naturally exhibit an overdensity at apoastron, because lower orbital velocities at this location make material spend a larger part of their orbital period there. At (sub)mm wavelengths, apocenter brightness due to this overdensity is expected to become predominant over that of the pericenter, which was indeed observed in the debris disc of Fomalhaut with ALMA observations at 1.3 mm \citep{2017ApJ...842....8M}.
As the ALMA/Atacama Compact Array (ACA) performs (sub)mm observations with resolutions similar to that of \textit{Herschel}/PACS, we proposed to resolve the double-lobed feature around \zR with the ALMA/ACA in both Band 6 and Band 7, that is, at 1.3mm and 870 microns, respectively. Our goal was to reveal this apocenter-glow and study the evolution of the lobes' brightness asymmetry from one wavelength to another, as it provides crucial information on the debris disc dust grains population \citep{2016ApJ...832...81P}. However, because of \textit{Herschel}'s large beam size, the possibility of background contamination of \textit{Herschel}'s observations is by no mean negligible, which led \citet{2018MNRAS.475.3046S} to suggest that the lobes seen in \textit{Herschel}/PACS observations are likely background objects. In addition, a final reduction of the \textit{Spitzer}/MIPS data by \citet{2014ApJ...785...33S} found there was no infrared excess at 24 microns. Therefore, our ALMA observations, which were carried out nearly 8 years after the \textit{Herschel} ones, were also potentially expected to assess critically the debris disc interpretation of the \textit{Herschel}/PACS observations. Indeed, as \zR exhibits a high proper motion, the double lobe structure would retain its position relative to the star if it were a debris disc, while background objects would instead retain their absolute position.

\section{Observations}\label{sec:obs}

We present here ALMA/ACA observations of \zR in Band 6  (230 GHz, 1.3mm), that were carried out from Nov 4th to Nov 18th 2017, and in Band 7 (345 GHz, 870 microns), that were carried out from Oct 27th to Nov 22th 2017, both under the project 2017.1.00786.S (PI: V. Faramaz).

Our Band 7 and Band 6 data comprised 8 and 9 separate observations, respectively, for which we summarize the characteristics in Table \ref{tab:obslogB7} and \ref{tab:obslogB6}. These were both taken using baselines ranging from 8.9 to 48.9 m. In Band 7, these correspond to angular scales of 20".14 and 3".67, respectively. Given the distance of the star (12 pc), this means that the spatial scales that were probed ranged from 44 to 242 AU. In Band 6, these baselines correspond to angular scales of 30".21 and 5".5, respectively, which in turn translates into probed spatial scales ranging from 66 to 363 AU.

The spectral setup in Band 7 consisted of three 2 GHz-wide spectral windows, centred on 334, 336, and 348 GHz, each one divided into 128 channels of width 15.625 MHz $(\sim 14\,\mathrm{km.s^{-1}})$. Although we did not expect primordial gas to be present in a system as old as \zR, we nevertheless used the fourth spectral window to probe CO gas, via the J=3-2 emission line, as CO can be released from collisions among planetesimals. Therefore, this 1 GHz-wide spectral window was centred on 346 GHz, and set with a larger number (2048) of finer channels, leading to a spectral resolution of 0.5 MHz $(\sim 0.5\,\mathrm{km.s^{-1}})$.  Following the same strategy, the spectral setup in Band 6 consisted of four spectral windows, each 2 GHz-wide, with three of them centred on 232.5, 245.5, and 247.5 GHz, and divided into 128 channels of width 15.625 MHz $(\sim 20\,\mathrm{km.s^{-1}})$, while the fourth spectral window was set to search for CO gas via the J=2-1 emission line and was centred on 230.5 GHz, with a larger number (2048) of finer channels, leading to a spectral resolution of 1 MHz $(\sim 1.3\,\mathrm{km.s^{-1}})$.

Since in Band 7, the primary beam has a diameter of 41" and the presumed apocenter of the debris disc surrounding \zR on \textit{Herschel} images is located approximately 15" from the star, we expected it to fall outside the central third of the primary beam, where sensitivity decreases. Therefore, we requested mosaic observations, and chose to set a pointing offset by 11".2 from the star along the major axis in the direction of the presumed apocenter of the disc, and set another pointing towards the presumed pericenter, while adjusting the distance between both pointings to be half of the primary beam Half Power Beamwidth (HPBW). The total on-source time was 6 h in Band 7.
In the case of our Band 6 observations, the primary beam has a diameter of 62" and the presumed apocenter of the debris disc surrounding \zR was expected to fall within the central third of the primary beam. Therefore there was no need for mosaic observations and we requested single-pointing observations. The total on-source time was 6.8 h in Band 6.

\begin{figure*}
\makebox[\textwidth]{\includegraphics[scale=0.37]{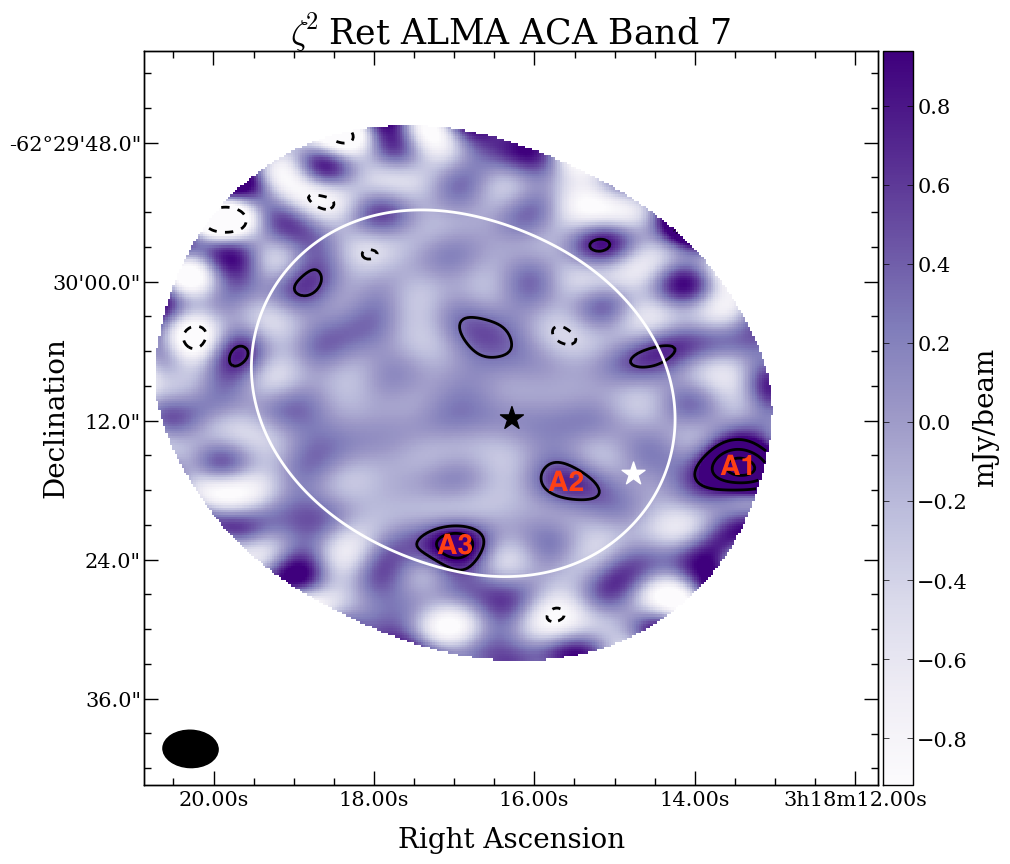}
\includegraphics[scale=0.37]{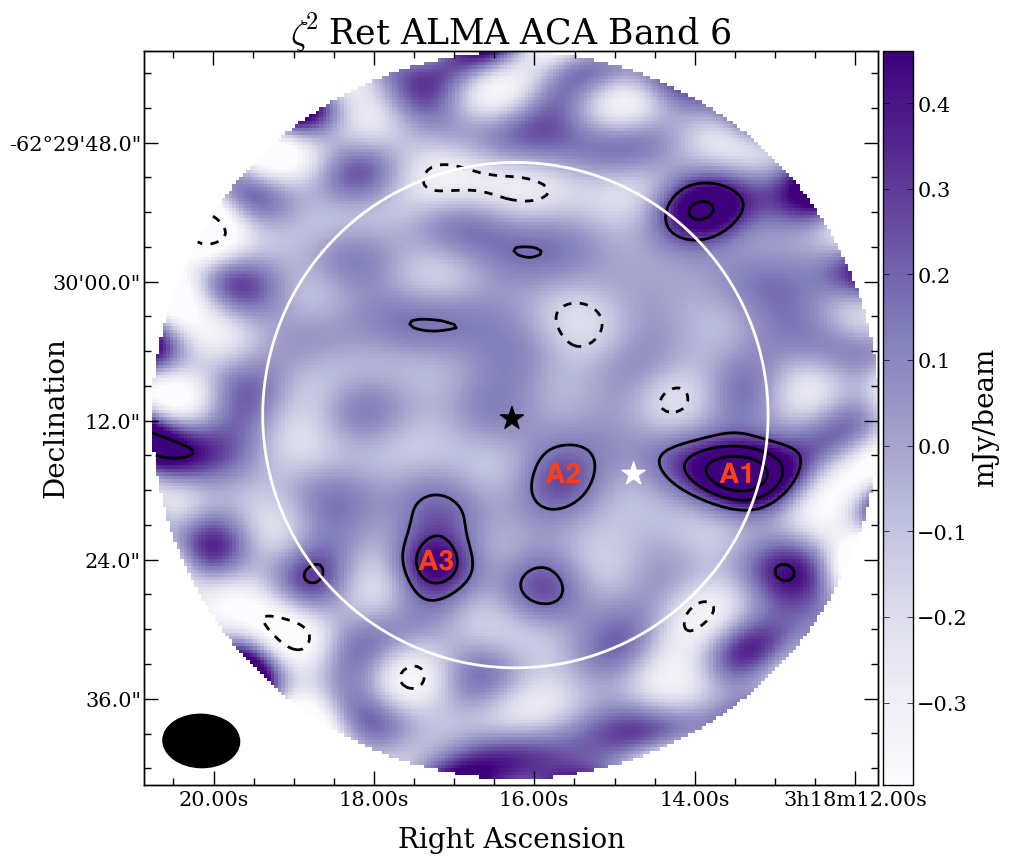}}
\caption[]{\emph{Left:} ALMA/ACA 870 microns continuum observations of \zR. Contours show the $\pm$2, 4, 6,... $\sigma$ significance levels, with $\sigma=150\,\mu\mathrm{Jy/beam}$. The synthesized beam, shown on the lower left side of the image, has dimensions $4".77 \times 3".2$, with position angle $88^{\circ}$. \emph{Right:} ALMA/ACA 1.3 continuum observations of \zR. Contours show the $\pm$2, 4, 6,... $\sigma$ significance levels, with $\sigma=66\,\mu\mathrm{Jy/beam}$. The synthesized beam, shown on the lower left side of the image, has dimensions $6".62 \times 4".60$, with position angle $88^{\circ}$. In both images, the position of the star is marked with a black star, while its position at the time of \textit{Herschel}/PACS observations is marked with a white star. The white line indicates the 50\% response of the primary beam, and the colorbars show the fluxes in mJy per beam.}
\label{fig:HR1010_continuum}
\end{figure*}

The data were calibrated using the pipeline provided by ALMA. We used the \emph{TCLEAN} algorithm and task in CASA version 5.1.1 \citep{2007ASPC..376..127M} to perform the image reconstruction of the continuum emission, that is, to obtain the inverse Fourier transform of the observed visibilities. In both cases, we combined the four spectral windows in order to recover the maximum signal-to-noise (S/N) ratio, and used a Briggs weighting scheme with robustness parameter set to 0.5. We further corrected the images for the primary beam, and show the results in Figure \ref{fig:HR1010_continuum}, where the position of \zR is marked with a black star. 
For our Band 7 image, we used a cell size of 0".2, and an image size of $286\times 286$ pixels in order to cover the primary beam. The resulting synthesized beam has dimensions $4".77 \times 3".2$, with position angle $88^{\circ}$. The rms was measured in a large region far from the sources present in the field of view, and was found to be $\sigma=150\,\mu\mathrm{Jy/beam}$.
For our Band 6 image, we used a cell size of 0".3, and an image size of $210\times 210$ pixels. The resulting synthesized beam has dimensions $6".62 \times 4".60$, with position angle $88^{\circ}$, while the rms was found to be $\sigma=66\,\mu\mathrm{Jy/beam}$.
 
\begin{figure*}
\makebox[\textwidth]{\includegraphics[scale=0.4]{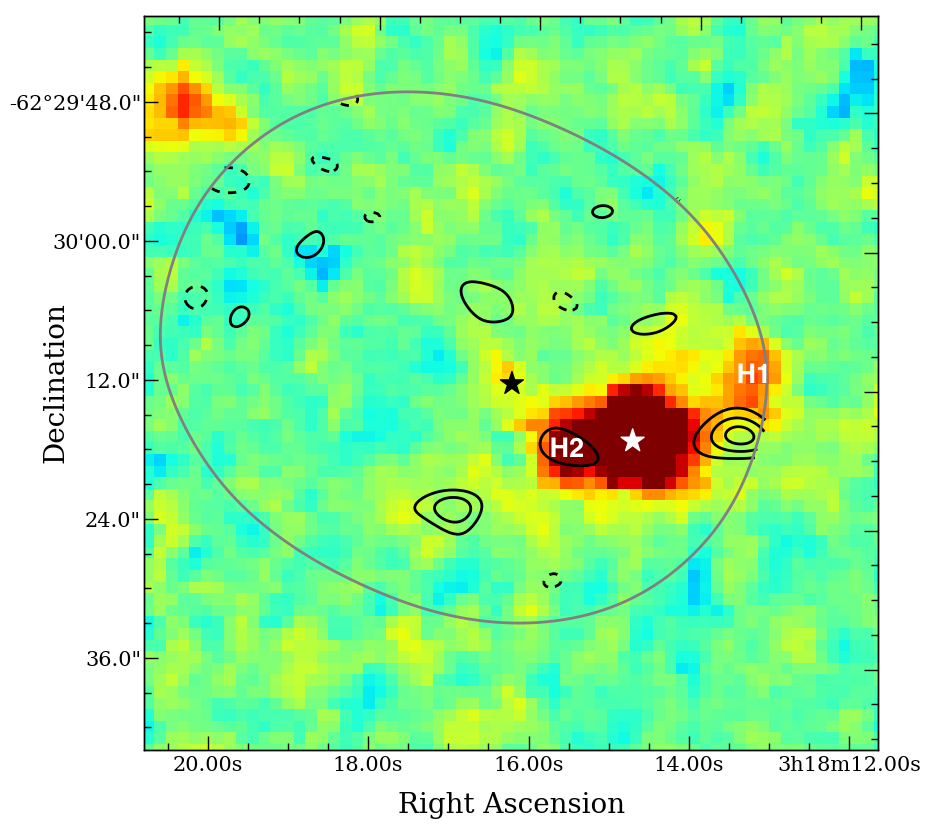}
\includegraphics[scale=0.4]{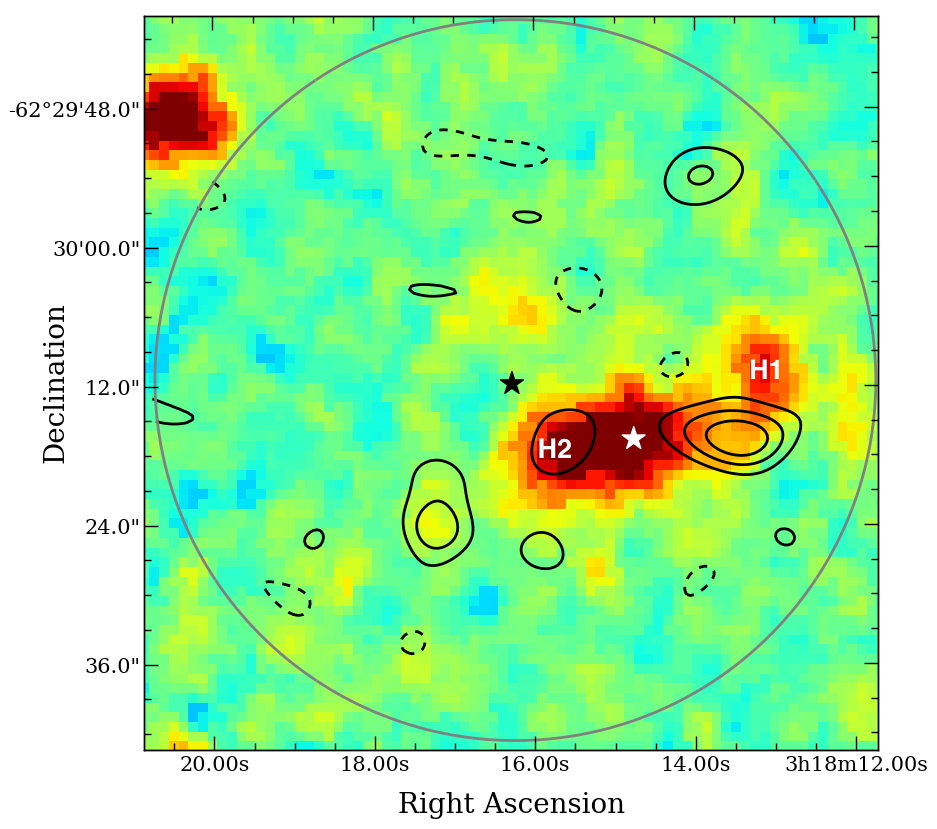}}
\caption[]{Comparison of the double lobe structure position between \textit{Herschel}/PACS and ALMA/ACA observations. \textit{Left:} ALMA/ACA 870 microns $\pm$2, 4, 6,... $\sigma$ significance levels contours (with $\sigma=150\,\mu\mathrm{Jy/beam}$), overlayed on \textit{Herschel}/PACS colormap at 70 microns. \textit{Right: } ALMA/ACA 1.3 mm $\pm$2, 4, 6,... $\sigma$ significance levels contours (with $\sigma=66\,\mu\mathrm{Jy/beam}$), overlayed on \textit{Herschel}/PACS colormap at 100 microns. In both images, the position of the star in ALMA observations is marked with a black star, while its position at the time of \textit{Herschel}/PACS observations is marked with a white star. The grey line shows the limits of the ALMA images.}
\label{fig:comp_Herschel_ALMA}
\end{figure*}

\begin{table*}

\begin{tabular*}{1.\textwidth}{@{\excs}ccccccccc}
\toprule
  Date  & Time & On source & $\mathrm{N_{Ant.}}$ & PWV & Avg Elev. & \multicolumn{3}{c}{Calibrators} \\[5pt]
 dd/mm/yyyy & (UTC) & (min) &   & (mm) & (deg) & Flux & Bandpass & Phase  \\[5pt]
\midrule
\midrule
27/10/2017 & 02:38:39.9 & 44.9 & 11  & 0.56-0.61 & 46.3 & J2258-2758 & J2258-2758 & J0303-6211  \\[2pt]
04/11/2017 & 05:15:49.7 & 44.9 & 11 & 0.48-0.50   & 48.2 & J0522-3627 & J0522-3627 & J0303-6211  \\[2pt]
09/11/2017 & 07:02:55.6  & 44.9 & 10 & 0.55-0.64  & 37.6 & J0522-3627 & J0522-3627 & J0303-6211  \\[2pt]
10/11/2017 & 06:24:53.3  & 44.9 & 10 & 1.06-1.20  & 41.1 & J0522-3627 & J0522-3627 & J0303-6211  \\[2pt]
11/11/2017 & 04:59:43.7 & 44.9 & 10 & 0.64-0.69 & 47.5 & J0522-3627 & J0522-3627 & J0303-6211  \\[2pt]
12/11/2017 & 04:56:19.4 & 44.9 & 10 & 0.56-0.61 & 47.5 & J0522-3627 & J0522-3627 & J0303-6211  \\[2pt]
19/11/2017 & 05:17:49.6 & 44.9 & 11  & 0.60-0.68 & 44.1 & J0522-3627 & J0522-3627 & J0516-6207  \\[2pt]
22/11/2017 & 04:22:05.9 & 44.9 & 11 & 0.53-0.57 & 47.2 & J0522-3627 & J0522-3627 & J0303-6211  \\[2pt]
\bottomrule     
\end{tabular*}
\captionof{table}{Summary of our ALMA/ACA observations at 870 microns (Band 7)}
\label{tab:obslogB7}

\begin{tabular*}{1.\textwidth}{@{\excs}ccccccccc}
\toprule
  Date  & Time & On source & $\mathrm{N_{Ant.}}$ & PWV & Avg Elev. & \multicolumn{3}{c}{Calibrators} \\[5pt]
 dd/mm/yyyy & (UTC) & (min) &   & (mm) & (deg) & Flux & Bandpass & Phase  \\[5pt]
\midrule
\midrule
04/11/2017 & 02:59:33.5 & 45.4 & 11  & 0.54-0.70 & 48.4 & J2258-2758 & J2258-2758 & J0303-6211  \\[2pt]
05/11/2017 & 06:32:44.0 & 45.4 & 10 & 1.11-1.28   & 43.7 & J0522-3627 & J0522-3627 & J0303-6211  \\[2pt]
06/11/2017 & 06:16:37.7  & 45.4 & 10 & 1.45-1.51  & 44.8 & J0522-3627 & J0522-3627 & J0303-6211  \\[2pt]
08/11/2017 & 05:17:42.0  & 45.4 & 10 & 1.06-1.20  & 48.2 & J0522-3627 & J0522-3627 & J0303-6211  \\[2pt]
08/11/2017 & 06:41:54.7 & 45.4 & 10  & 1.10-1.20 & 41.9 & J0522-3627 & J0522-3627 & J0303-6211  \\[2pt]
11/11/2017 & 06:46:43.8 & 45.4 & 10 & 0.56-0.61 & 40.3 & J0522-3627 & J0522-3627 & J0303-6211 \\[2pt]
15/11/2017 & 04:58:05.7 & 45.4 & 11 & 1.82-2.35 & 47.8 & J0522-3627 & J0522-3627 & J0303-6211  \\[2pt]
16/11/2017 & 04:32:52.9 & 45.4 & 11  & 0.64-0.69 & 48.9 & J0522-3627 & J0522-3627 & J0303-6211  \\[2pt]
18/11/2017 & 05:21:43.9 & 45.4 & 10 & 1.09-1.12 & 44.9 & J0522-3627 & J0522-3627 & J0516-6207  \\[2pt]
\bottomrule     
\end{tabular*}
\captionof{table}{Summary of our ALMA/ACA observations at 1.3 mm (Band 6)}
\label{tab:obslogB6}

\end{table*}

\section{Results}\label{results}

We recover a double-lobe structure similar to that seen in \textit{Herschel}/PACS data both in Band 6 and Band 7, and with the presumed apocenter West lobe being brighter than the East lobe. Nevertheless, it appears that \zR is not surrounded by this double-lobe structure.

\zR possesses a high proper motion of $1331.15 \pm 0.36\,\mathrm{mas/year}$ in Right Ascension and $648.52 \pm 0.43\,\mathrm{mas/year}$ in Declination \citep{2016A&A...595A...1G,2018arXiv180409365G}. Therefore, it is expected to have displaced by 11".5 in the North-East direction (position angle of $64^{\circ}$) since it was observed with \textit{Herschel} in February 2010, that is, 7.75 yr before our ALMA observations. We show the position of \zR at the time of \textit{Herschel} observations with a white star in Figure \ref{fig:HR1010_continuum}. We find that this position falls in the middle of the double-lobe structure visible both in Band 7 and Band 6, and closer to the East lobe (A2) than to the West one (A1), which strikingly resembles the configuration seen in \textit{Herschel} observations. We further show a comparison of this double-lobe structure position seen in our ALMA observations with that seen in \textit{Herschel} observations in Figure \ref{fig:comp_Herschel_ALMA}. 

The A1 lobe is found to have a peak brightness of $\sim 1\,\mathrm{mJy/beam}$ at 870 microns, with an integrated flux of $\sim 1\,\mathrm{mJy}$, and is unresolved. However, these measurements should be taken with great caution, as A1 is far from the phasecentre of the ALMA image at this wavelength, where uncertainties are very large and fitting difficult. This is not the case at 1.3 mm, where A1 is found to have a peak brightness of $\sim 565\pm 47\,\mu\mathrm{Jy/beam}$, with an integrated flux of $794 \pm 106\,\mu\mathrm{Jy}$, and is found to be marginally resolved. We modeled it with a Gaussian ellipse, which once deconvolved from the beam, is found to have major axis $5".83 \pm 1".53$, minor axis $1".54 \pm 0".79$, and position angle $87 \pm 12^\circ$. 

A1 has a position that is $\sim 5"$ in the South East direction compared to the West lobe seen in \textit{Herschel}/PACS images (labeled H1 in Figure \ref{fig:comp_Herschel_ALMA}). Consequently, if A1 and H1 emission comes from the same source, it possesses a proper motion and therefore, is not an extragalactic source. However, it does not share proper motion with \zR and is then unrelated to it. On the other hand, sources seen at \textit{Herschel} wavelengths do not necessarily possess counterpart emission at ALMA wavelengths, and vice-versa. For instance, the source labeled A3 in Figure \ref{fig:HR1010_continuum} and seen at ALMA wavelengths, does not show far-IR counterpart emission in \textit{Herschel} observations, as can be seen in Figure \ref{fig:comp_Herschel_ALMA}. Therefore, it is not impossible either that A1 and H1 emissions are unrelated, with the H1 having no emission at ALMA wavelengths, and A1  having no emission at \textit{Herschel} wavelengths. 

The A2 lobe seen in our ALMA observations is marginally resolved in Band 7, and fitted by a Gaussian ellipse which once deconvolved from the beam, has a major axis $5".11 \pm 1".33$, minor axis $1".19 \pm 0".92$, and a position angle $63\pm15^ \circ$. The larger beam size in Band 6 does not allow to resolve this source. We measure a peak brightness of $735\pm 75\,\mu\mathrm{Jy/beam}$ at 870 microns, with an integrated flux of $1.20\pm 0.19 \,\mathrm{mJy}$, and a peak brightness of $336\pm 21 \,\mu\mathrm{Jy/beam}$ at 1.3 mm, with an integrated flux of $511\pm50\,\mu\mathrm{Jy}$. It shares common location with the East lobe H2 seen in \textit{Herschel} observations, and therefore, if both emissions come from the same object, then it is an extragalactic source. However, it is not excluded either that these two emissions are unrelated. 

Overall, the possibility for far-IR sources to be too faint for ALMA to detect, and ALMA sources to be too faint for \textit{Herschel} to detect does not allow us to tell for certainty  whether the A1 and H1, and A2 and H2 emissions are related or not. 
It also means that the double-lobe structure seen with \textit{Herschel}/PACS really could be  a debris disc, which is too faint at ALMA wavelengths to be detected. 
We test this hypothesis extrapolating the flux of such a disc from \textit{Herschel}/SPIRE photometric measurements at 250 microns, adopting a spectral index of 2.6. This choice was made based on statistics of the spectral indices derived from debris discs SED modelling for which SCUBA-2 photometry at 850 microns was available, and which show that the majority of debris discs have a spectral index of 2.6 or smaller \citep{2017MNRAS.470.3606H}. Assuming that the disc surface brightness is uniform, and that such a disc is an extended structure which requires 4 ALMA-ACA beams in Band 6 and 6 beams in Band 7 to be covered, it should have been detected in our observations at SNR of at least 2.5 in Band 7 and at least 3 in Band 6. In addition, the surface brightness of such a hypothesized eccentric $(e \gtrsim 0.3)$ disc is not expected to be uniform: the pericenter and apocenter of the disc are expected to be the brightest portions of the disc, with the apocenter being in addition up to 30\% brighter than the pericenter due to the apocenter-glow phenomenon at these wavelengths \citep[see Equation (8) of][]{2016ApJ...832...81P}. Hence the SNR quoted above are very conservative lower limits, and we should have detected such a disc in our observations if it truly existed. Therefore, we exclude this scenario.

Finally, no CO line emission was detected in Band 6, nor in Band 7.

\section{Conclusions}\label{sec:conclu}

The position of the A2 lobe seen with ALMA coincides with the H2 lobe seen with \textit{Herschel}. If these two emissions are related, then it is stationary and most likely a background source, but they might as well not be related. It is also difficult to say whether the A1 lobe seen with ALMA and the H1 lobe seen with \textit{Herschel} are related emissions. If they are, then it is a nearby source that has displaced South East since \textit{Herschel} observations, and thus is not co-moving with \zR.  If they are not related, then the source H1 has become too faint at ALMA wavelengths while A1 appeared South East to the \textit{Herschel} location of H1. 

Nevertheless, we can exclude the possibility that the double-lobe feature seen with \textit{Herschel} really is a debris disc, as even a very faint counterpart (sub-)mm emission should have been detected at the sensitivities achieved in our ALMA observations. Therefore, our ALMA/ACA observations show without ambiguity that the double lobe structure seen in \textit{Herschel} observations was not the signature of an eccentric debris disc.

Revisiting the \textit{Herschel}/PACS photometry and SED reported in \citet{2010A&A...518L.131E} in the light of our findings, then \zR possesses no detectable infrared excess above its expected photospheric flux. These results are in accordance with those of \citet{2018MNRAS.475.3046S}, who found no significant infrared excess to \zR at \textit{Herschel}/PACS wavelengths, and thus concluded that the features seen near this star are the result of background confusion.

As the photometric measurement of \textit{Herschel}/PACS at 70 microns is in accordance with that of \textit{Spitzer}, we further conclude that \textit{Spitzer} observations at 70 microns were contaminated as well.

In conjunction with our ALMA data, and since \zR has been observed with sensitivities of $\sigma=66\,\mathrm{\mu Jy/beam}$ and $\sigma=150\,\mathrm{\mu Jy/beam}$ in Band 6 and Band 7, respectively, we conclude that if there is any debris disc around \zR, then its  surface brightness falls below the $3\sigma$ detection limits, that is, it does not exceed $5.7\,\mathrm{\mu Jy/arcsec^2}$ at 1.3 mm, and $26\,\mathrm{\mu Jy/arcsec^2}$ at 870 microns. It would be one and two orders of magnitude dimmer than the well-studied discs around HD 202628 (Faramaz et al., in prep.) and Fomalhaut \citep{2017ApJ...842....8M}, respectively.

Although these observations report a non-detection around \zR, this study overall demonstrates that using the ALMA/ACA can be a powerful technique to clarify the effects of background confusion for \textit{Herschel}-detected debris discs. In particular, peculiar \textit{Herschel}/PACS observations of infrared excesses peaking at 160 microns among the DUNES sample have been interpreted to be "cold debris discs" \citep{2011A&A...536L...4E,2013ApJ...772...32K}, whereas \citet{2014ApJ...784...33G} showed that this hypothesis was indistinguishable from background confusion. Therefore, we advocate the use of ALMA-ACA to disentagle these two hypothesis.

\section*{Acknowledgements}

This paper makes use of the following ALMA data: ADS/JAO.ALMA\#2017.1.00786.S. ALMA is a partnership of ESO (representing its member states), NSF (USA) and NINS (Japan), together with NRC (Canada), MOST and ASIAA (Taiwan), and KASI (Republic of Korea), in cooperation with the Republic of Chile. The Joint ALMA Observatory is operated by ESO, AUI/NRAO and NAOJ. The National Radio Astronomy Observatory is a facility of the National Science Foundation operated under cooperative agreement by Associated Universities, Inc. . 
VF's postdoctoral fellowship is supported by the Exoplanet Science Initiative at the Jet Propulsion Laboratory, California Inst. of Technology, under a contract with the National Aeronautics and Space Administration.
MB acknowledges support from the Deutsche Forschungsgemeinschaft (DFG) through project Kr 2164/15-1. 
A.~B., J,~C., and J.~O. acknowledge financial support from the ICM (Iniciativa Cient\'ifica Milenio) via the N\'ucleo Milenio de Formaci\'on Planetaria grant. J.~O acknowledges financial support from the Universidad de Valpara\'iso, and from Fondecyt (grant 1180395).
This work has made use of data from the European Space Agency (ESA) mission {\it Gaia} (\url{https://www.cosmos.esa.int/gaia}), processed by the {\it Gaia} Data Processing and Analysis Consortium (DPAC, \url{https://www.cosmos.esa.int/web/gaia/dpac/consortium}). Funding for the DPAC has been provided by national institutions, in particular the institutions participating in the {\it Gaia} Multilateral Agreement.

\vspace{1cm}

Copyright 2018. All rights reserved.




\bibliographystyle{mnras}
\bibliography{biblio_HR1010} 



%
%


\bsp	
\label{lastpage}
\end{document}